\documentclass[mathleft
]{an}
\usepackage{graphicx}
\usepackage{times}
\begin{document}

\Pagespan{1}{}
\Yearpublication{2007}%
\Yearsubmission{2006}%
\Month{12}%
\Volume{328}%
\Issue{3--4}%

\title{A near-infrared/optical/X-ray survey in the centre of $\sigma$~Orionis} 

\author{Jos\'e A. Caballero\inst{1,2}\fnmsep\thanks{Alexander von Humboldt
Fellow at the MPIA.}
}
\titlerunning{A survey in the centre of $\sigma$~Orionis}
\authorrunning{J.A. Caballero}
\institute{
Max-Planck-Institut f\"ur Astronomie, K\"onigstuhl 17, D-69117 Heidelberg,
Germany, \email{caballero@mpia.de} 
\and 
Instituto de Astrof\'{\i}sica de Canarias, E-38205 La Laguna, Tenerife, Spain}

\received{Dec 2006}
\accepted{7 May 2007}
\publonline{later}


\abstract{%
Because of the intense brightness of the OB-type multiple star system
$\sigma$~Ori, the low-mass stellar and substellar populations close to the
centre of the very young $\sigma$~Orionis cluster is poorly know. 
I present an $IJHK_{\rm s}$ survey in the cluster centre, able to detect from
the massive early-type stars down to cluster members below the deuterium burning
mass limit.  
The near-infrared and optical data have been complemented with X-ray imaging. 
Ten objects have been found for the first time to display high-energy emission.
Previously known stars with clear spectroscopic youth indicators and/or X-ray
emission define a clear sequence in the $I$ vs. $I-K_{\rm s}$ diagram.
I have found six new candidate cluster members that follow this sequence.
One of them, in the magnitude interval of the brown dwarfs in the cluster,
displays X-ray emission and a very red $J-K_{\rm s}$ colour, indicative of a
disc.
Other three low-mass stars have excesses in the $K_{\rm s}$ band as well.
The frequency of X-ray emitters in the area is 80$\pm$20\,\%.
The spatial density of stars is very high, of up to 1.6$\pm$0.1\,arcmin$^{-2}$. 
There is no indication of lower abundance of substellar objects in the cluster
centre.
Finally, I also report two cluster stars with X-ray emission located
at only 8000--11000\,AU to $\sigma$~Ori AB, two sources with peculiar colours
and an object with X-ray emission and near-infrared magnitudes similar to
those of previously-known substellar objects in the cluster.} 

\maketitle

\section{Introduction}
\label{intro}

The star $\sigma$~Ori, with a visual magnitude of $V \approx$ 3.8\,mag, is the 
fourth brightest star in the Orion Belt and the brightest star of the
$\sigma$~Orionis cluster, to which it gives the name\footnote{I am using the
name $\sigma$~Ori for the star and the name $\sigma$~Orionis for the cluster. 
Given the brightness of $\sigma$~Ori, the star may have been in Gan's Treatise
on Stars (Gan De approx. 350 BC), Hipparchus catalogue (Hipparchus 135 BC),
Almagest (Ptolemy approx. 150), Book of Fixed Stars (al Sufi 964) and
Zij-i Sultani star catalogue (Ulugh Beg 1437). 
It also appears, among others, in Uranometria (Bayer 1603), Flamsteed's
catalogue (Flamsteed 1712), Bonner Sternverzeichniss (Schoenfeld 1886) and Henry
Draper Catalogue (Cannon \& Pickering 1925).
The first spectroscopic study of $\sigma$~Ori as a pair was performed by Frost
\& Adams (1904).}.  
Nowadays it is thought to be at least a quintuple system of OB-type stars that
illuminates the mane of the \object{Horsehead Nebula} and injects turbulence and
high-energy radiation to the intra-cluster medium.
The central $\sigma$~Ori~AB pair consists of an O9.5V primary and a B05.V
secondary separated $\sim$0.25\,arcsec.
It is one of the shortest-period ($P$ = 158\,a) and most massive visual O-type
visual binaries known (Heintz 1974, 1997; Hartkopf 1996; Mason et al. 1998). 
The combined mass of \object{$\sigma$~Ori~A} and \object{$\sigma$~Ori~B} at the
most probable heliocentric distance of the cluster is $\sim$35\,M$_\odot$.  
The components \object{$\sigma$~Ori~C} (A0V, M = 2.7$\pm$0.4\,M$_\odot$) and
\object{$\sigma$~Ori~D} (B2V, M = 6.8$^{+1.8}_{-1.2}$\,M$_\odot$) are located
11\,arcsec to the southwest and 13\,arcsec to the east of $\sigma$~Ori AB,
respectively.  
The fifth component of the system, \object{$\sigma$~Ori E} (B2Vp, M = 
7.4$^{+1.5}_{-1.4}$\,M$_\odot$), is an helium-rich peculiar star, photometric
variable and strong X-ray emitter. 
It is at 42\,arcsec to the east-northeast of the primary, at a project physical
separation of about 15\,000\,AU (the masses of $\sigma$~Ori C, D and E have been
taken from Caballero 2007). 

The above-mentioned injection of turbulence and energy may be the origin of the
wealthy population of brown dwarfs and isolated planetary-mass objects below the
deuterium burning mass limit in the $\sigma$~Orionis cluster (B\'ejar et al.
1999, 2001, 2004b; Zapatero Osorio et al. 2000; Caballero et al. 2004, 2007;
Scholz \& Eisl\"offel 2004; Kenyon et al. 2005; Gonz\'alez-Garc\'{\i}a et al.
2006).   
The cluster, located in the eastern part of the \object{Ori OB 1 b} Association,
is very young (3$\pm$2\,Ma), nearby (360$^{+70}_{-60}$\,pc) and free of
extinction if compared to other rich star forming regions like \object{Taurus},
\object{Chamaeleon I} or \object{$\rho$ Ophiuchi}.
A summary of the basic properties of $\sigma$~Orionis was given in
Caballero (2007). 

\begin{figure*}
\includegraphics[height=0.32\textwidth]{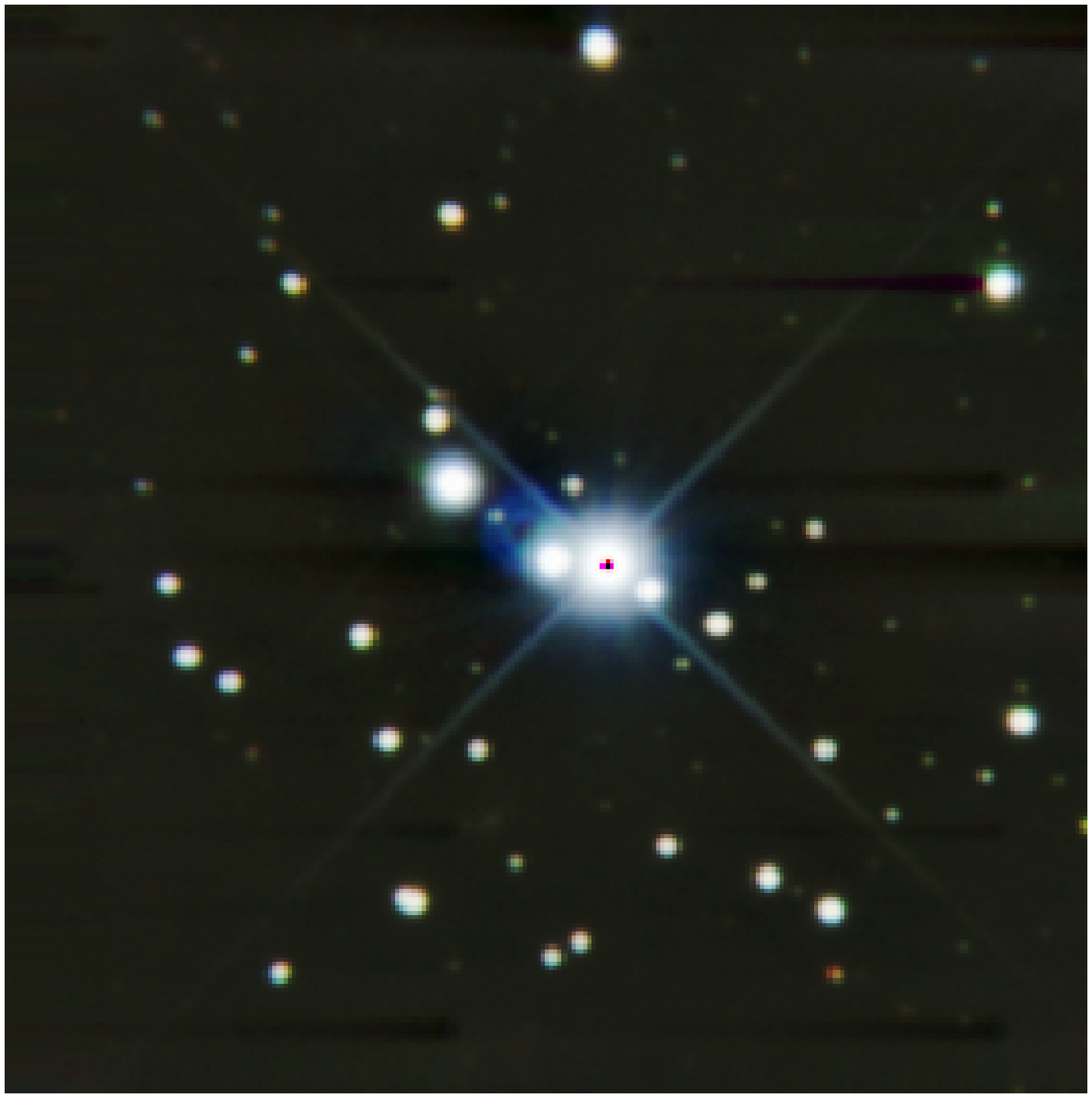}
\includegraphics[height=0.32\textwidth]{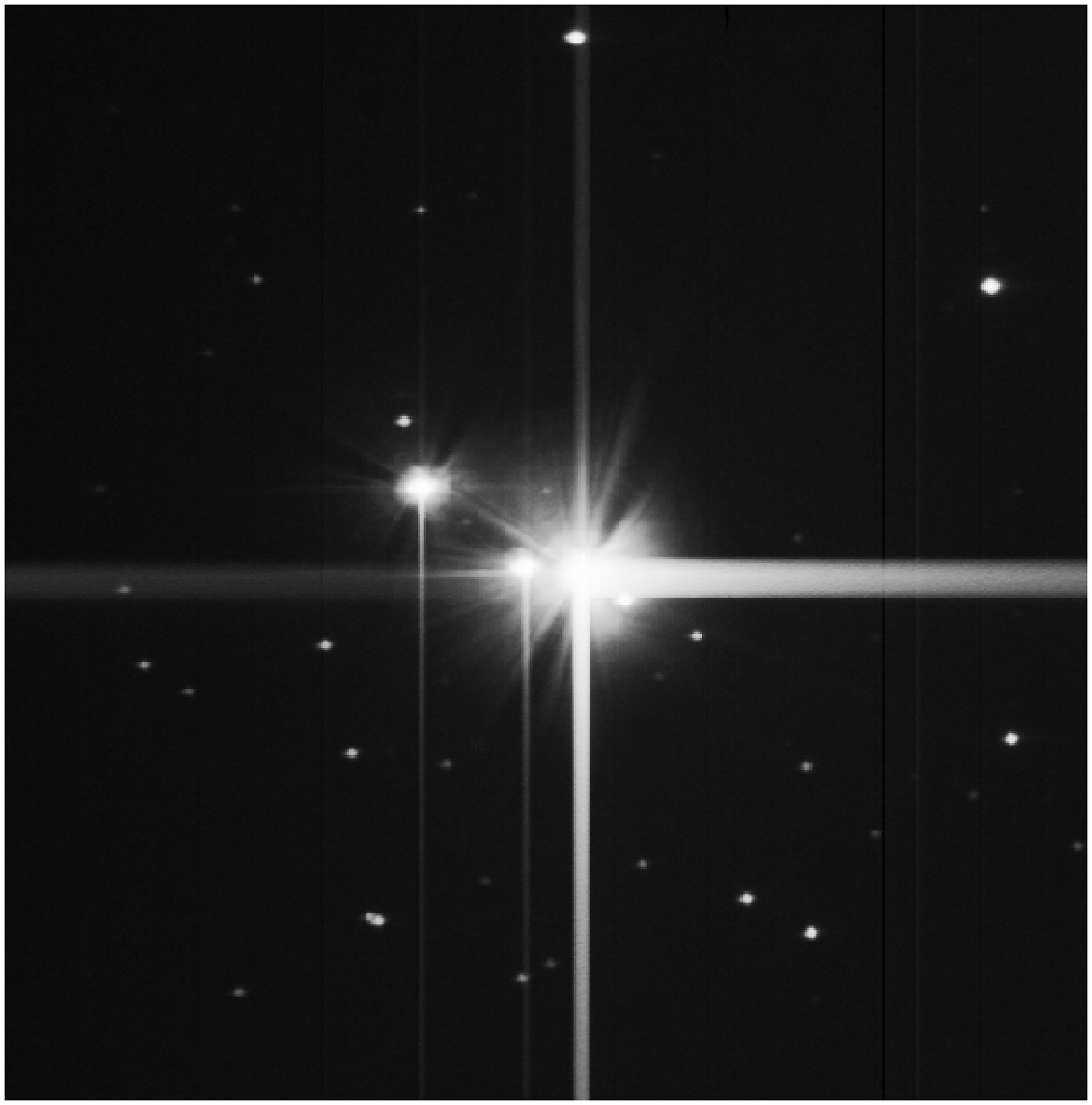}
\includegraphics[height=0.32\textwidth]{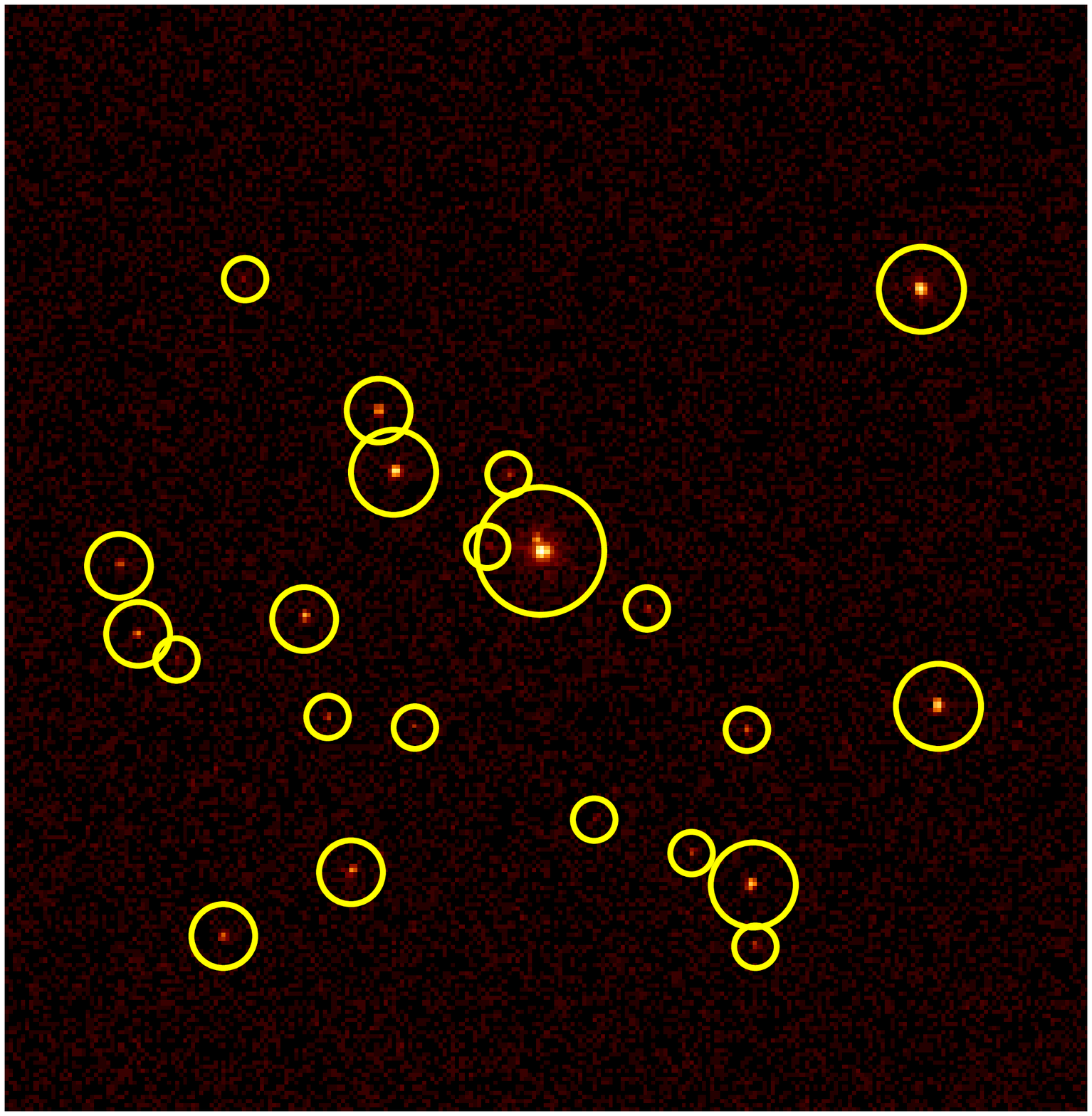}
\caption{Images at different wavelengths of the optical/near-infrared/X-ray
survey in the centre of the $\sigma$~Orionis cluster.
The sizes are about 4.3 $\times$ 4.3\,arcmin$^2$.
North is up and east is left.
{\em Left window:} $JHK_{\rm s}$ CAIN-II/TCS false-colour composite image (blue
is for $J$, green is for $H$ and $K_{\rm s}$ is for red).
{\em Middle window:} central part of the $I$ CCD/IAC-80 final image.
Note the bleeding lines due to saturation of $\sigma$~Ori.
{\em Right window:} central part of the X-ray HRC-I/{\em Chandra} Space
Telescope image.
X-ray sources are encircled.
Radii of the circles roughly indicate the strength of the X-ray emission.
Colour versions of all the Figures in this paper are available electronically.
}
\label{F01}
\end{figure*}

The glare in the optical caused by the tremendous luminosity of the quintuple
early-type system is the reason for the relative poor knowledge of the
low-mass star and substellar populations near the core of the cluster. 
The innermost known candidate brown dwarf and planetary-mass object are
\object{S\,Ori 71} and \object{S\,Ori 68}, at $\sim$4 and $\sim$8\,arcmin to
$\sigma$~Ori AB, respectively.
At a separation less than $\sim$2.5\,arcmin, there are only seven known
stellar members with spectroscopy, eight photometric cluster candidates, and
several X-ray sources (some of them without known optical counterpart), apart
for the five members of the OB star system  
(Wolk 1996; Oliveira, Jeffries \& van Loon 2004; Sherry, Walter \& Wolk 2004;
Weaver \& Babcock 2004; Sanz-Forcada, Franciosini \& Pallavicini 2004; Kenyon et
al. 2005; Burningham et al. 2005; Franciosini et al. 2006).    
However, the faintest cluster member known within the 2.5-arcmin radius is more
than 1\,mag brighter than the hydrogen burning limit at $I \sim$ 16.5, $J \sim$
14.5\,mag, that marks the boundary between stars and brown dwarfs in the
$\sigma$~Orionis cluster (Caballero et al. 2007). 
Therefore, all the substellar objects and many faint star cluster members close
to the centre may have eluded detection by current surveys, mostly based only
on standard broad-band optical imaging. 

Different techniques have been recently applied to search for low-mass cluster
members close to the $\sigma$~Ori system.
Using mid-infrared imaging (8--18\,$\mu$m) with TIMMI-2 at the 3.6\,m ESO
Telescope, van Loon \& Oliveira (2003) detected \object{$\sigma$~Ori~IRS1} at
only 3.3\,arcsec to the north-east north of $\sigma$~Ori~AB.
Its nature is still under debate, since it is a relatively strong X-ray emitter
(Sanz-Forcada et al. 2004) and their $J$ and $H$ magnitudes from adaptive-optics
imaging locate the source in the stellar domain (Caballero 2005).
The latter found in the same adaptive-optics near-infrared search another star
with X-ray emission at $\sim$21\,arcsec north of $\sigma$~Ori AB.
Sherry, Walter \& Wolk (2005), using $VI$-band quasi-speckle imaging in the
0.5\,arcmin-radius central region, also detected this source (labeled {\em star
2}) and a new candidate cluster member ({\em star 1}, in the direction to
$\sigma$~Ori~E). 
The first brown dwarf to be detected in the area still remains kept out of
sight.

The aim of this work is to study the core of the $\sigma$~Orionis cluster with
near-infrared imaging, complemented with red optical and X-ray data, and to
search for new low-mass stellar and brown-dwarf cluster members at relatively
close physical separations ($<$ 40\,000\,AU) to the central OB star system.
The frequency of substellar objects compared to stars in a highly energetic
environment like this can provide further constraint to ultra-low-mass
formation scenarios.

\section{Near-infrared and optical imaging}

\begin{table}
\centering
\caption{Completeness and limiting magnitudes of the near-infrared and optical
images.}  
\label{ijhks}
\begin{tabular}{lcc}
\hline
Band 		& Completeness 	& Limiting	\\
 		& (mag) 	& (mag)		\\
\hline
$I$ 		& 17.5 		& 18.5	\\
$J$ 		& 17.5 		& 19.0	\\
$H$ 		& 17.5 		& 18.5	\\
$K_{\rm s}$	& 16.5 		& 17.5	\\
\hline
\end{tabular}
\end{table}

\subsection{$JHK_{\rm s}$ CAIN-II/Telescopio Carlos S\'anchez}

I used the near-infrared instrument CAIN-II (Cabrera-Lavers et al. 2006)
attached to the 1.5\,m Telescopio Carlos S\'anchez (TCS) at the Observatorio del
Teide to obtain series of $JHK_{\rm s}$ images centred in the quintuple $\sigma$
Ori star system. 
The detector of CAIN-II is a NICMOS-3 with 256 $\times$ 256 pixels of
1.00\,arcsec size, providing a field of view of $\sim$4.3 $\times$
4.3\,arcmin$^2$. 
The core of the cluster was observed in 16 ($J$), 8 ($H$) and 18 ($K_{\rm s}$)
occasions during eight nights between 2002 Sep 27 and 2004 Oct 08.
Every visit comprised 5\,min exposure imaging in the cluster centre using
a 5-point dithering pattern. 
Individual exposure times per frame ranged from 6\,s ($K_{\rm s}$) to 10\,s
($J$ and $H$).
All the nights were clear, but the seeing varied between 1.0 and 2.0\,arcsec
from one campaign to other.
The images were aligned and combined to get three final images, one for each
pass-band.
A false-colour image after combining the three near-infrared final images is
shown in the left panel of Fig. \ref{F01}.

\begin{table*}
\centering
\caption{Objects with optical spectroscopy in the survey area.}
\label{spektra}
\begin{tabular}{lcccr}
\hline
Name 				& Sp. Type	& Remarks 				& Other name(s) 	& Refs. \\ 
\hline
$\sigma$~Ori AB+IRS1$^a$	& O9.5V+	& OBA, triple 				& 48 Ori 		& Morgan et al. (1955); Edwards (1976) \\ 
$\sigma$~Ori C  		& A2V		& OBA 					& BD--02 1326C 		& \\ 
$\sigma$~Ori D  		& B2V		& OBA 					& HD 37468D 		& \\ 
$\sigma$~Ori E  		& B2Vp		& OBA, He rich 				& V1030 Ori 		& Lesh (1968) \\ 
\object{GSC 04771--01147}  	& K0V 		& Li {\sc i}, H$\alpha$, SB 		& 4771--1047		& Wolk (1996) \\ 
\object{rJ053838--0236}   	& K8V		& Li {\sc i}, H$\alpha$, RV		& [OJV2004] 7 		& Zapatero Osorio et al. (2002) \\ 
\object{rJ053841--0237}   	& K3V		& Li {\sc i}, strong H$\alpha$ 		& \object{[WB2004] 22} 	& Wolk (1996); Weaver \& Babcock (2004) \\ 
\object{R053847--0237}$^b$   	& K5V		& Li {\sc i}, strong H$\alpha$, double 	& \object{SWW 102}+\object{SWW 149}	& Wolk (1996) \\ %
\object{rJ053851--0236}   	& K5V		& Li {\sc i}, strong H$\alpha$, double  & 			& Wolk (1996) \\ %
\object{[KJN2005] 8}   		& 		& Li {\sc i}, RV, low $g$  		& 			& Kenyon et al. (2005) \\ %
\object{B 3.01--67}   		& 		& RV, low $g$  				& 			& Burningham et al. (2005) \\ %
\hline
\end{tabular}
	\begin{itemize}
	\item[$^a$] Hierarchical triple system described in the text.
	\item[$^b$] Not in SIMBAD (I follow the original nomenclature by Wolk 
	1996).
	\end{itemize}
\end{table*}

I performed PSF photometry on the final images to get the instrumental
magnitudes in the $J$, $H$ and $K_{\rm s}$ bands of 98 sources (there are only
59 2MASS point-like sources in the same 4.3 $\times$ 4.3\,arcmin$^2$ area).
Their number of counts in the peak ranged between the non-linear limit of the
detector of CAIN-II and five times the standard deviation of the background.
The reduction and photometric analysis of the data were done with standard
tasks within the IRAF environment.
I used stars in common with 2MASS to obtain the coordinates and calibrated
magnitudes of the sources in the TCS data. 
Uncertainties in the zero points varied between 0.060 ($H$) and 0.090\,mag
($K_{\rm s}$). 
The completeness magnitudes of the final images, shown in Table \ref{ijhks},
were $\sim$2\,mag fainter than in the 2MASS catalogue.
Besides, there is an appreciable number of detected sources fainter than the
completeness limit.

The three brightest stars in the field of view, $\sigma$~Ori AB, D and E, are
brighter than $J =$ 7.2\,mag and saturate in the TCS images.
The $J$-band PSF photometry of another two relatively bright stars is affected
by their closeness to the border of the field of view. 
For the five of them, I have used the magnitudes tabulated by 2MASS.
Thanks to this incorporation, there is information in this search for objects
with $J$-band magnitudes between 4.8 and $\sim$20\,mag (i.e. covering all the
spectral domain in the cluster from O9.5 V + B0.5V down to intermediate L).
The near-infrared source $\sigma$~Ori IRS1 and the $\sigma$~Ori AB pair were
not resolved neither by TCS nor by 2MASS.

\subsection{$I$ CCD/Telescopio IAC-80}

On 2004 Nov 17, I obtained $I$-band imaging of the cluster centre with the CCD
camera at the 0.8\,m Telescopio IAC-80, also at the Observatorio del Teide.
The (old) CCD camera had a detector 1k $\times$ 1k Thomson with a pixel size of
0.4325\,arcsec\,px$^{-1}$.
I took 30 images of 30\,s in the field of view of 7.4 $\times$ 7.4\,arcmin$^2$. 
I also took advantage of the sky being covered by a thin layer of cirrus,
which avoided the strong saturation of the $\sigma$~Ori stars and the subsequent
formation of bleeding lines in the detector.
Since the brightest stars were in the non-linear regime in the 30\,s exposures, 
very short $VRI$ exposures were also taken.
Images were again reduced (over-scan and flat-field corrected), aligned and
combined using common IRAF tasks.
The central part of the final combined $I$-band image, overlapping with the
$JHK_{\rm s}$ survey, is shown in the middle panel of Fig. \ref{F01}. 

The PSF photometry on the final deep optical image was performed in the same way
as the near-infrared ones.
The photometric calibration was done using 21 stars in common to the work by
Wolk (1996), who also used a broad-band $I$ Johnson filter.
The error in the zero point of the optical images was 0.090\,mag.
The magnitudes of $\sigma$~Ori C, D and E were extracted from the short-time
$I$-band image, after scaling with the deep one (the combined magnitude of
$\sigma$~Ori AB, saturating at any exposure time, was taken from the Hipparcos
Catalogue). 
The calibrated photometry was also compared to that offered by the DENIS
catalogue (Epchtein et al. 1997).
Discarding two objects whose photometry is affected by high optical
background due to bleeding lines, the standard deviation between the calibrated
IAC-80 and DENIS photometry is 0.11\,mag. 
The completeness and limiting magnitudes are given in Table \ref{ijhks}.
Due to the meteorological restriction, the exposure time (15\,min in total) and
the observational procedure (short individual exposure times lead stronger
contribution of the noise readout), only the cluster members brighter than $J
\sim$ 16.5\,mag have an $I$-band counterpart.

\section{Analysis}

\subsection{Known objects in the survey area}

\subsubsection{Stars with optical spectroscopy}
\label{S_spektra}

Table \ref{spektra} shows the basic spectroscopic data of the components of the
$\sigma$~Ori system and the remaining seven stars with optical spectroscopy.
All of them display some kind of feature associated to extreme youth: early
spectral type (OBA), Li {\sc i} $\lambda$6707.8\,\AA~in absorption, H$\alpha$
$\lambda$6562.8\,\AA~in (strong) emission and/or low gravity spectroscopic
features (from the abnormal pseudo-equivalent widths, pEW, of some alkali
absorption lines if compared with field dwarfs of the same spectral type).
The stars marked with ``strong H$\alpha$'' are accretors according to the
criterion defined by Barrado y Navascu\'es \& Mart\'{\i}n (2003). 
The measured pEW(Li {\sc i})s vary between +0.22 and +0.53\,\AA~which, for their
respective spectral types, correspond to primordial abundance of lithium.
Some of them have radial velocity measurements consistent with membership
in the cluster (see Caballero 2007 for a compilation of radial velocities
measured in $\sigma$~Orionis). 
Multiplicity (double, triple, spectroscopic binary --SB--) and
spectroscopic peculiarity (helium rich) are also indicated in the Table.

Sanz-Forcada et al. (2004) tabulated the spectral types of GSC 04771--01147 and
rJ053841--0237 at K0 and K3, as well.
Their spectral types estimated from X-ray data obtained with the European Photon
Imaging Cameras (EPIC) at the {\em XMM-Newton} Space Telescope are consistent
with the determinations from optical spectra (J. Sanz-Forcada, priv. comm.).

I will consider the eleven stars as spectroscopically confirmed cluster members.

\subsubsection{Photometric cluster member candidates}
\label{S_photometric}

Apart from the known stars with optical spectroscopy, there are also eight
photometric cluster member candidates without spectroscopic information.
They were detected in the $(B)VRI$ searches by Wolk (1996) and Sherry et al.
(2004).
Some of them were also followed-up in the $JHK_{\rm s}$ bands.
The original names of the eight stars are given in Table \ref{photometric}.
The objects found by Wolk (1996) were classified as pre-main sequence stars.

Besides, Sherry et al. (2005) presented the source {\em star 1} as a new
candidate cluster member.
It is not shown in the Table and will be discussed later.
They considered the source {\em star 2} as a very red foreground M dwarf.

\begin{table}
\centering
\caption{Known photometric cluster member candidates in the survey area.}
\label{photometric}
\begin{tabular}{lcr}
\hline
Name 				& Other name(s) 	& Refs.$^a$ \\ 
\hline
\object{P053844--0233}  	&			& 1 	\\ 
\object{P053842--0237}  	& SWW 48		& 1, 2 	\\ 
\object{SWW 78}  		& 			& 2 	\\ 
\object{SWW 35}  		& 			& 2 	\\ 
\object{P053847--0234}  	& SWW 29		& 1, 2 	\\ 
\object{P053843--0237}  	& SWW 15		& 1, 2 	\\ 
\object{P053841--0236}  	&			& 1 	\\ 
\object{P053850--0237}  	& SWW 18		& 1, 2 	\\ 
\hline
\end{tabular}
	\begin{itemize}
	\item[$^a$] (1): Wolk (1996); (2): Sherry et al. (2004).
	They are not in SIMBAD (I follow the original nomenclature by Wolk 
	1996 and Franciosini et al. 2006).
	\end{itemize}
\end{table}

\subsubsection{X-ray sources}
\label{S_xray}

Stars and substellar objects in young ($<$10\,Ma) star-forming regions like
$\sigma$~Orionis are known to display magnetic activity, mostly detected through
X-ray observations (Simon, Herbig \& Boesgaard 1985; Walter et al. 1988;
Neuh\"auser et al. 1995; Neuh\"auser \& Comer\'on 1998; Webb et al. 1999
-- see a review on high-energy processes in young stellar objects in Feigelson
\& Montmerle 1999).  
In the area under study, all the {\em ROSAT} X-ray sources identified in the
seminal work by Wolk (1996) were studied in detail by Franciosini et
al. (2006). 
They limited the analysis to the {\em XMM-Newton} EPIC 0.3--7.8\,keV energy
band. 
The four brightest X-ray sources in the area ($\sigma$~Ori AB, $\sigma$~Ori E,
GSC 04771--01147 and rJ053841--0237) were also spectroscopically investigated
with EPIC (Sanz-Forcada et al. 2004; $\sigma$~Ori AB was studied with the
Reflection Grating Spectrometer at a higher resolution as well). 
According to Pallavicini, Franciosini \& Randich (2004), rJ053841--0237 shows
evidence of rotational modulation of its X-ray emission with period and
amplitude of $\sim$8.5\,h and $\sim$25\,\%.
It could be due to the heterogeneous distribution of active regions on the
stellar surface.
Finally, it has been reported the detection of strong flares from the
hot B2Vp star $\sigma$~Ori~E (Pallavicini, Sanz-Forcada \& Franciosini 2002;
Groote \& Schmidt 2004; Sanz-Forcada et al. 2004).
It has been suggested that the flares are originated from an unseen late-type
companion.

\begin{table}
\centering
\caption{X-ray sources from Franciosini et al. (2006) (with NX designation) and
undetected late-type cluster members or candidates (without NX designation) in
the survey area.}  
\label{xray}
\begin{tabular}{lccc}
\hline
Name 				& NX 	& Count rate 		& W96 	\\ 
 			& {\em (XMM)} 	& (10$^{-3}$\,s$^{-1}$) & {\em (ROSAT)}	\\ 
\hline
$\sigma$~Ori AB  		& 80	& 440    $\pm$ 2    	& yes	\\ 
$\sigma$~Ori E  		& 84	& 199.4  $\pm$ 1.7  	& yes	\\ 
GSC 04771--01147  		& 65	& 138.8  $\pm$ 1.4  	& yes	\\ 
rJ053838--0236  		& 64	&  37.6  $\pm$ 0.7  	& yes	\\ %
rJ053841--0237  		& 70	& 107.2  $\pm$ 1.2  	& yes	\\ 
R053847--0237  			& 88	&   4.0  $\pm$ 0.3  	& yes	\\ %
P053842--0237  			& --	&  $<$ 63.9  		&	\\ %
SWW 78  			& 87	&   5.9  $\pm$ 0.6  	&	\\ %
\object{2MASS J05384828--0236409} &  91 &   0.74 $\pm$ 0.14 	&	\\ %
SWW 35  			& 92	&  13.6  $\pm$ 0.5  	&	\\ %
rJ053851--0236  		& 102	&   5.0  $\pm$ 0.3  	& yes	\\ %
P053847--0234  			& --	&  $<$ 0.35  		&	\\ %
P053843--0237  			& --	&  $<$ 0.51  		&	\\ %
\object{2MASS J05385173--0236033} & 103 &   3.5  $\pm$ 0.2  	&	\\ %
P053841--0236  			& 71	&   1.08 $\pm$ 0.17 	&	\\ %
P053850--0237  			& 98	&   3.7  $\pm$ 0.2  	&	\\ %
KJN2005 8  			& --	&  $<$ 4.98  		&	\\ %
B 3.01--67  			& --	&  $<$ 0.53  		&	\\ %
\object{NX 77} 			&  83   &   0.63 $\pm$ 0.18 	&	\\ 
\object{NX 99} 			&  99   &   0.58 $\pm$ 0.12 	&	\\ 
\hline
\end{tabular}
\end{table}

Table \ref{xray} shows the names of the X-ray sources from Franciosini et al.
(2006) in the near-infrared survey and the identification (NX) and count
rates in that paper.
The last column indicates if the X-ray source was also detected with {\em ROSAT}
by Wolk (1996). 
In the list, 11 stars are spectroscopically confirmed cluster members or
known photometric cluster member candidates.
The stars NX~103 and NX~91, with significance of detection of 28.2 and 6.6,
are X-ray sources identified with possible cluster candidates from 2MASS (Table
1 in Franciosini et al. 2006).
Finally, NX~77 and NX~99 are unidentified X-ray sources with no known
counterpart within 5\,arcsec (significance of detection 5.6 and 7.7,
respectively).
Both of them were detected in EPIC MOS1 and MOS2 independent cameras, which
supports the reliability of the detections.

Additionally, Franciosini et al. (2006) gave 3$\sigma$ upper limits for five
late-type cluster members or candidates (shown in the Table \ref{xray} without
NX designation).  
For two of them, especially in the case of P053842--0237, the upper limits are
not quite restrictive.

\subsection{X-ray HRC+ACIS/{\em Chandra} Space Telescope}
\label{S_newxray}

I have downloaded X-ray public data from the {\it Chandra} Data Archive to
search for new high-energy counterparts of the near-infrared sources in my
survey. 
In particular, I have studied the central regions of the images obtained by
Adams et al. (2004) with the High Resolution Camera, HRC-I, and by Skinner et
al. (2004) with the Advanced CCD Imaging Spectrometer, ACIS.
Both instruments are attached to the {\it Chandra} Space Telescope. 
The HRC-I image is deeper than the ACIS one, and has an excellent spatial
resolution (see right panel of Fig. \ref{F01}).
Both of them allow to resolve the $\sigma$~Ori AB + $\sigma$~Ori IRS1 system
(see Section \ref{MM}). 

\begin{table}
\centering
\caption{New X-ray sources from {\em Chandra} data in the survey area.}
\label{newxray}
\begin{tabular}{lcc}
\hline
Name 				& ACIS 	& HRC-I \\ 
\hline
$\sigma$~Ori D  		& 	& yes	\\ 
P053842--0237  			& 	& yes	\\ 
P053843--0237  			& 	& yes	\\ 
\object{2MASS J05384301--0236145}$^a$ &  & yes?	\\ 
\object{2MASS J05384970--0234526}$^b$ &  & yes?	\\ 
KJN2005 8  			& 	& yes	\\ 
B 3.01--67  			& 	& yes	\\ 
\object{Mayrit 21023} 		& yes 	& yes	\\ 
\object{2MASS J05384146--0235523}$^c$ &  & yes?	\\ 
\object{2MASS J05384123--0237377}$^d$ & yes & yes	\\ 
\hline
\end{tabular}
	\begin{itemize}
	\item[$^a$] Mayrit 30241. 
	\item[$^b$] Mayrit 100048. 
	\item[$^c$] Mayrit 50279. 
	\item[$^d$] Mayrit 11128. 
	\end{itemize}
\end{table}

I have identified the {\it Chandra} counterparts of 14 of the 15 X-ray
sources from Franciosini et al. (2006) in Table \ref{xray} (all of them
except NX 77). 
Besides, I have detected for the first time the faint X-ray counterparts (in the
HRC-I image) of ten near-infrared sources, shown in Table \ref{newxray}.
The significance of the detections is larger than 3, except for three 2MASS
stars that are detected at the $\sim$2$\sigma$ level (marked with ``yes?'').
Two faint near-infrared sources were also detected in the ACIS image.
For four of the candidate cluster members with new X-ray counterpart,
Franciosini et al. (2006) had given upper limits on the X-ray emission from EPIC
data (one case is P053842--0237, without a restrictive upper limit). 
The analysis of the X-ray data corresponding to the new sources (count
rates, accurate detection significance, energy) will be given in a forthcoming
paper (Caballero \& Sanz-Forcada, in prep.).  
No X-ray source without near-infrared counterpart has been additionally found.

Among the ten new X-ray sources, there are a B2V star, two confirmed low-mass
cluster members (with low gravity spectroscopic features), two previously-known
photometric cluster member candidates, the near-infrared source found at
$\sim$21\,arcmin to $\sigma$~Ori~AB by Caballero (2005) (Mayrit 21023; see
Section \ref{intro} for the object discovery and Section \ref{selection} for its
designation) and four 2MASS sources. 
There are TCS and IAC-80 data for all of them.

\subsection{Selection of candidate cluster members}
\label{selection}

\begin{table*}
\centering
\caption{CAIN-II near-infrared and IAC-80 optical photometry of selected
candidate cluster members with $IJHK_{\rm s}$ photometry.}
\label{photon}
\begin{tabular}{l c c c c c c c}
\hline
Identification 		& $\alpha$ 		& $\delta$		& $I$	 	& $J$ 		 & $H$		  & $K_{\rm s}$	    & Name 			\\
 			& (J2000) 		& (J2000)		& (mag)	 	& (mag)	 	 & (mag)	  & (mag)	    & 				\\
\hline
        Mayrit AB$^a$	& 05 38 44.76		& $-$02 36 00.2   	& 4.02$\pm$0.01	&  4.75$\pm$0.26 &  4.64$\pm$0.25 & 4.490$\pm$0.016 & $\sigma$~Ori AB		\\ 
\object{Mayrit  41062}	& 05 38 47.20 		& $-$02 35 40.5		& 6.86$\pm$0.10	&  6.97$\pm$0.03 &  6.95$\pm$0.03 &  6.95$\pm$0.03  & $\sigma$~Ori E		\\ 
\object{Mayrit  13084}	& 05 38 45.62 		& $-$02 35 58.9   	& 6.98$\pm$0.03	&  7.12$\pm$0.03 &  7.22$\pm$0.03 &  7.26$\pm$0.02  & $\sigma$~Ori D		\\ 
\object{Mayrit  11238}	& 05 38 44.12 		& $-$02 36 06.3    	& 9.14$\pm$0.10	&  9.22$\pm$0.09 &  9.24$\pm$0.08 &  9.15$\pm$0.09  & $\sigma$~Ori C		\\ 
\object{Mayrit 114305}	& 05 38 38.49 		& $-$02 34 55.0 	&10.88$\pm$0.09 &  9.98$\pm$0.08 &  9.26$\pm$0.08 &  9.16$\pm$0.10  & GSC 04771--01147		\\ 
\object{Mayrit 123000}	& 05 38 44.80		& $-$02 33 57.6     	&11.12$\pm$0.09 &  9.64$\pm$0.09 &  9.41$\pm$0.09 &  9.13$\pm$0.13  & P053844--0233		\\ 
\object{Mayrit 105249}	& 05 38 38.23 		& $-$02 36 38.4     	&12.41$\pm$0.09 & 11.47$\pm$0.10 & 10.46$\pm$0.06 & 10.39$\pm$0.10  & rJ053838--0236		\\ 
\object{Mayrit  97212}	& 05 38 41.29 		& $-$02 37 22.6 	&12.71$\pm$0.09 & 11.48$\pm$0.08 & 10.74$\pm$0.06 & 10.67$\pm$0.09  & rJ053841--0237		\\ 
\object{Mayrit  92149}  & 05 38 47.92 		& $-$02 37 19.2      	&13.39$\pm$0.09 & 11.65$\pm$0.11 & 10.89$\pm$0.10 & 10.52$\pm$0.12  & R053847--0237		\\ 
\object{Mayrit  83207}	& 05 38 42.28 		& $-$02 37 14.8 	&13.04$\pm$0.09 & 11.73$\pm$0.08 & 10.98$\pm$0.06 & 10.81$\pm$0.09  & P053842--0237		\\ 
\object{Mayrit  53049}	& 05 38 47.46 		& $-$02 35 25.2 	&13.14$\pm$0.09 & 11.84$\pm$0.08 & 11.00$\pm$0.09 & 10.65$\pm$0.09  & SWW 78			\\ 
\object{Mayrit  30241}	& 05 38 43.02 		& $-$02 36 14.6 	&13.31$\pm$0.09 & 11.91$\pm$0.08 & 11.03$\pm$0.06 & 10.62$\pm$0.09  & 				\\ 
\object{Mayrit  67128}	& 05 38 48.29 		& $-$02 36 41.0 	&13.86$\pm$0.09 & 11.99$\pm$0.08 & 11.33$\pm$0.06 & 11.08$\pm$0.09  & NX 91			\\ 
\object{Mayrit  61105}	& 05 38 48.68 		& $-$02 36 16.2 	&13.72$\pm$0.09 & 12.14$\pm$0.08 & 11.41$\pm$0.06 & 11.09$\pm$0.09  & SWW 35			\\ 
\object{Mayrit 102101}	& 05 38 51.45 		& $-$02 36 20.6     	&14.36$\pm$0.09 & 12.44$\pm$0.03 & 11.77$\pm$0.07 & 11.52$\pm$0.10  & rJ053851--0236		\\ 
\object{Mayrit  91024}	& 05 38 47.19 		& $-$02 34 36.8 	&14.52$\pm$0.09 & 12.66$\pm$0.08 & 11.79$\pm$0.08 & 11.20$\pm$0.10  & P053847--0234		\\ 
\object{Mayrit  68191}	& 05 38 43.87 		& $-$02 37 06.8 	&14.82$\pm$0.09 & 12.85$\pm$0.08 & 12.11$\pm$0.06 & 11.71$\pm$0.09  & P053843--0237		\\ 
\object{Mayrit 105092}	& 05 38 51.74 		& $-$02 36 03.4      	& 14.7$\pm$0.5  & 12.91$\pm$0.03 & 12.27$\pm$0.07 & 11.94$\pm$0.10  & NX 103			\\ 
\object{Mayrit 100048}	& 05 38 49.70		& $-$02 34 52.6 	&14.81$\pm$0.09 & 12.98$\pm$0.08 & 12.32$\pm$0.08 & 12.01$\pm$0.10  & 				\\ 
\object{Mayrit  68229}	& 05 38 41.36 		& $-$02 36 44.5 	&14.50$\pm$0.09 & 13.00$\pm$0.08 & 12.27$\pm$0.06 & 12.10$\pm$0.09  & P053841--0236		\\ 
\object{Mayrit 124140}	& 05 38 50.03 		& $-$02 37 35.5 	&14.97$\pm$0.10 & 13.03$\pm$0.08 & 12.36$\pm$0.06 & 12.12$\pm$0.09  & P053850--0237		\\ 
\object{Mayrit  94106}	& 05 38 50.78 		& $-$02 36 26.8     	&15.13$\pm$0.09 & 13.13$\pm$0.08 & 12.48$\pm$0.06 & 12.18$\pm$0.09  & [KJN2005] 8		\\ 
\object{Mayrit  53144}	& 05 38 46.85 		& $-$02 36 43.5 	&15.23$\pm$0.09 & 13.21$\pm$0.08 & 12.59$\pm$0.06 & 12.32$\pm$0.09  & B 3.01--67		\\ 
\object{Mayrit  89175}	& 05 38 45.28 		& $-$02 37 29.3 	&15.46$\pm$0.09 & 13.30$\pm$0.08 & 12.60$\pm$0.06 & 12.14$\pm$0.09  &				\\ 
\object{Mayrit  21023}	& 05 38 45.31 		& $-$02 35 41.3 	&14.7$\pm$0.5$^b$&13.41$\pm$0.09 & 12.98$\pm$0.06 & 12.73$\pm$0.09  & {\em star 2}		\\ 
\object{Mayrit  50279}	& 05 38 41.46 		& $-$02 35 52.3 	&16.15$\pm$0.09 & 13.95$\pm$0.08 & 13.26$\pm$0.06 & 12.91$\pm$0.09  &				\\ 
\object{Mayrit  36273}	& 05 38 42.39 		& $-$02 36 04.4 	&15.7$\pm$0.5$^b$&14.23$\pm$0.08 & 13.56$\pm$0.06 & 13.22$\pm$0.09  &				\\ 
\object{Mayrit 111208}	& 05 38 41.24		& $-$02 37 37.7 	&17.55$\pm$0.11 & 16.09$\pm$0.09 & 14.91$\pm$0.06 & 13.87$\pm$0.09  &				\\ 
\hline
\end{tabular}
	\begin{itemize}
	\item[$^a$] The objects \object{Mayrit A} ($\sigma$~Ori A),
	\object{Mayrit B} ($\sigma$~Ori B) and \object{Mayrit 3022} ($\sigma$
	Ori IRS1) are not resolved. 
	\item[$^b$] $I$-band photometry taken from DENIS (the stars are below 
	strong bleeding lines or the glare caused by $\sigma$~Ori AB in the
	IAC-80 image).  
	\end{itemize}
\end{table*}

\begin{figure}
\includegraphics[width=0.47\textwidth]{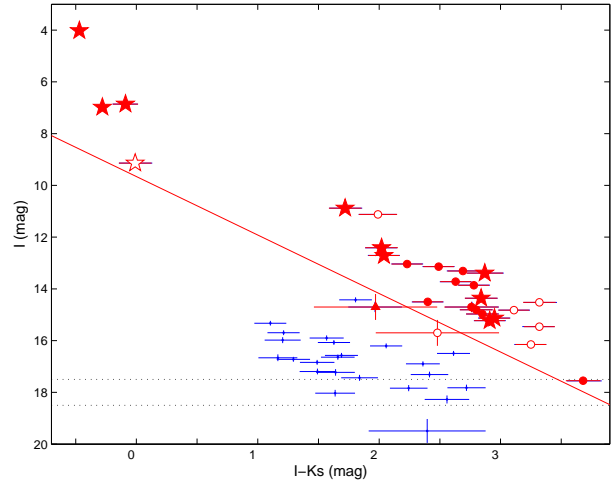}
\caption{$I$ vs. $I-K_{\rm s}$ colour-magnitude diagram of all the sources found
in the survey area.
Spectroscopically confirmed and photometric candidate cluster members are shown
with stars and circles, respectively.
Filled symbols are for objects with X-ray counterpart.
Probable fore- and background objects are marked with small points.
The triangle marks Mayrit 21023 (see text).
The solid line indicates the selection criterion, while the dotted lines mark
the completeness and limiting magnitudes of the optical search.
In the electronic version, cluster members and candidates are drawn in red and
contaminants are drawn in blue.}
\label{F02}
\end{figure}

I have correlated the near-infrared sources detected in the TCS survey with the
optical sources found in the IAC-80 image, the 2MASS objects in the area, the
stars with optical spectroscopy in Section \ref{S_spektra}, the photometric
cluster member candidates in Section \ref{S_photometric} and the X-ray sources
in Sections \ref{S_xray} and \ref{S_newxray}. 
There are $IJHK_{\rm s}$ data for almost 90 sources, which allow to study their
membership in the cluster from colour-magnitude diagrams. 
Only about a dozen faint near-infrared sources, $J >$ 16.5\,mag, have no optical
counterpart.

The 11 spectroscopically confirmed $\sigma$~Orionis members define a clear
sequence in the $I$ vs. $I-K_{\rm s}$ colour-magnitude diagram in Fig.
\ref{F02}. 
The sequence matches with that of the X-ray sources (except for star Mayrit
21023, with a large error in the optical photometry).
This fact strongly supports the hypothesis of all the stars with X-ray
emission in this survey being young objects that belong to the $\sigma$~Orionis
cluster.
I have sketched in the diagram a straight line that is $\sim$0.2\,mag blue-wards
of the lower envelope of the confirmed cluster members and X-ray emitters.
Red-wards of the line, there are 26 stars, including the 11 confirmed cluster
stars, the eight previously-known photometric cluster member candidates and 22
out of the 25 X-ray sources. 
There is no optical counterpart for the faint sources NX 77 and NX 99 (I have
not detected the near-infrared counterpart of NX 77 at all).
The remaining ``blue'' X-ray emitter is Mayrit 21023, which optical photometry
is strongly affected by the glare of $\sigma$~Ori~AB.
However, accounting for the large photometric error bars, it may be located
red-wards of my selection criterion.
The same occurs with Mayrit 36273, although this star has not been found to
display X-ray emission.
I will also consider both stars as candidate cluster members.

The coordinates, $IJHK_{\rm s}$ photometry and names of the 28 selected
candidate cluster members are given in Table \ref{photon}. 
In the first column, I provide the ``Mayrit'' designation for the cluster
members and candidates found in this survey.
I have tried to avoid further confusion on the nomenclature of members in the
$\sigma$~Orionis cluster (like the use of the ``S\,Ori'' designation for cluster
members that are {\em not} associated to the emission-line star S Ori --HD
36090--; B\'ejar et al. 1999) by using the acronym Mayrit\footnote{
The acronym {\em Mayrit} comes from the Arabic {\em al-Majrit} ``source of
water'', from where the name of the city of Madrid evolved.  
This is a tribute to the Arabic Astronomy, which gave the names to
\object{Alnilam}, \object{Alnitak} and \object{Mintaka}, the other three bright
stars of the Orion Belt.} plus a running number.

\section{Results and discussion}

\subsection{New low-mass cluster members}
\label{new}

I have detected six new candidate cluster members with $J$-band
magnitudes in the range 11.9 to 16.1\,mag (they have a blank in the last column
in Table \ref{photon}).
Four of them are among the five faintest selected objects in the survey.
Three are also faint X-ray sources with low significance of detection in the
HRC-I image.
However, Mayrit 111208 displays reliable detections in both HRC-I and ACIS
images. 

The mass and effective temperature (T$_{\rm eff}$) of the brightest new
candidate cluster member, Mayrit 30241, are about 0.6\,M$_\odot$ and 3600\,K
using the theoretical models of Baraffe et al. (1998) and an age of 3\,Ma.
The masses and T$_{\rm eff}$ of the other new candidate cluster members
are below about 0.26\,M$_\odot$ and 3300\,K (see, however, 
discussions on uncertainties of theoretical models at very young ages in
Baraffe et al. 2002 and in Chabrier et al. 2005).
The faintest object, the X-ray emitter Mayrit 111208, will be discussed
later.

\subsection{$K_{\rm s}$ excesses, X rays and discs}
\label{discs}

\begin{figure}
\includegraphics[width=0.47\textwidth]{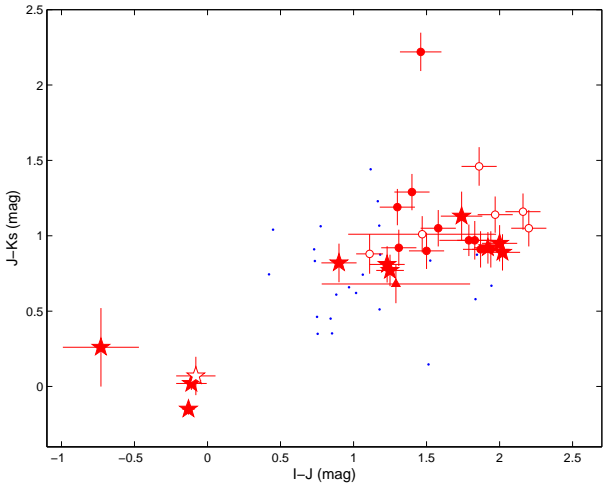}
\caption{The same as in Fig. \ref{F02}, but for the $J-K_{\rm s}$ vs. $I-J$
colour-colour diagram. 
Error bars of probable fore- and background contaminants are not drawn for
clarity.} 
\label{F03}
\end{figure}

There is evidence of circum(sub)stellar discs around four objects in Table
\ref{photon} from flux excess in the $K_{\rm s}$ band.
Their names and $I-J$ and $J-K_{\rm s}$ colours are given in Table
\ref{probablediscs}.
Their $J-K_{\rm s}$ colours are redder for their respective $I-J$ colours than
the rest of cluster members and candidates.
The colour-colour diagram in Fig. \ref{F03} illustrates the discussion on
near-infrared excesses.
In particular, the $J-K_{\rm s}$ colours of the four objects with probable discs
are redder than 1.15\,mag, and up to 2.22$\pm$0.13\,mag in the case of Mayrit
111208. 
This colour is extraordinary red, since there are only three known $\sigma$
Orionis members with $J-K_{\rm s} >$ 2.0\,mag from 2MASS data (\object{V505
Ori}, \object{V603 Ori} and \object{[WB2004] 10}). 
They are very active classical T Tauri stars with strong broadened H$\alpha$
emission or the sources of Herbig-Haro objects (Haro \& Moreno 1953;
Salukvadze 1987; Za\-pa\-te\-ro Osorio et al. 2002; Oliveira et al. 2004; Weaver
\& Babcock 2004). 
Mayrit 111208 is, however, even redder than them.
Its reddening is evident in the color version of left panel in Fig. \ref{F01}
(Mayrit 111208 is the red source close to the south-west corner).
In contrast, it displays an $I-J$ colour quite blue for its $J$ magnitude.
It would have not been selected as a probable cluster member from an $I$ vs.
$I-J$ colour-magnitude diagram using the lower envelope of spectroscopically
confirmed cluster members.
This ``blueing'' may be due to an optical veiling extending from the blue to the
red optical. 
This fact supports the use of the $I$ vs. $I-K_{\rm s}$ diagram for the
selection in this work.
Mayrit 111208 is close to the border of the TCS field of view and, hence, in an
area of lower exposure due to the dithering pattern. 
However, the DENIS $I$ and 2MASS $JHK_{\rm s}$ magnitudes reasonable agree (at
the 2$\sigma$ level) with the IAC-80 $I$ and TCS $JHK_{\rm s}$ magnitudes.
Mayrit 111208 is $\sim$1.6\,mag fainter than the hydrogen burning mass
limit in $\sigma$~Orionis (i.e. it may lie in the substellar domain).
If not a brown dwarf, then the central star should be obscured by at least
1.6\,mag in the $J$ band by a (probably edge-on) disc.
In this case, Mayrit 111208 would be by far the most extinguished star in
$\sigma$~Orionis (there is only one known high-mass star with a large
extinction, \object{IRAS 05358--0238}; Oliveira \& van Loon 2004).
This hypothesis is favoured by the very recent analysis of IRAC and MIPS
{\em Spitzer} Space Telescope data in the $\sigma$~Orionis cluster by
Hern\'andez et al. (2007), who classify Mayrit 111208 as a Class I object.

\begin{table}
\centering
\caption{Sources with probable discs in the survey area.}
\label{probablediscs}
\begin{tabular}{lccc}
\hline
Identification 	& $I-J$ 	& $J-K_{\rm s}$ & Name \\ 
	 	& (mag) 	& (mag)	 	& \\ 
\hline
Mayrit 53049	& 1.30$\pm$0.12	& 1.19$\pm$0.12	& SWW 78 \\ %
Mayrit 30241	& 1.40$\pm$0.12	& 1.29$\pm$0.12	& \\ %
Mayrit 91024	& 1.86$\pm$0.12	& 1.46$\pm$0.13	& P053847--0234 \\ %
Mayrit 111208	& 1.46$\pm$0.20	& 2.22$\pm$0.13	& \\ %
\hline
\end{tabular}
\end{table}

Among the other three objects with probable discs, there is a previously known
star with X-ray emission, ~ Mayrit 53049 (Sherry et al. 2004; Franciosini
et al. 2006). 
It is located at only 15.8\,arcsec ($\sim$5700\,AU) to the He-rich star Mayrit
41062 ($\sigma$~Ori E).
With a combined mass of $\sim$7.4 + 0.6\,M$_\odot$, they could be
gravitationally bound and may form a binary within a hierarchical $\sigma$~Ori
system. 
On the other hand, Mayrit 30241 is a previously-unknown star with a probable
X-ray emission (see Section \ref{MM}). 
The last star with a probable disc, Mayrit 91024 (Wolk 1996;
Sherry et al. 2004), is also very red ~ ($J-K_{\rm s}$ = 1.46$\pm$0.13\,mag) and
its spectral energy distribution from $I$ to $K_{\rm s}$ is similar to those of
other confirmed stars in $\sigma$~Orionis.
However, it is not an X-ray emitter.
Franciosini et al. (2006) imposed a restrictive upper limit of 0.00035 counts
per second in their EPIC observations.
I have not detected the X-ray counterpart of Mayrit 91024 in the HRC-I/ACIS
data, either.
Rather than attributing the lack of X-ray emission on a geometric factor or an
active/quiescent periodicity in a (probably weak-line) T Tauri star, I consider
that the high energy photons produced in the chromosphere are absorbed by
circumstellar material. 
This confirms the earlier results by Neuh\"auser et al. (1995), who founded
this effect on a large sample of weak-line and classical T Tauri stars in
Taurus.
Therefore, young naked low-mass stars and high-mass brown dwarfs would be easier
to detect in X rays.
This scenario is supported by the fact that the two known brown dwarfs in
$\sigma$~Orionis with X-ray emission do not have appreciable flux excess at
8.0\,$\mu$m (Sanz-Forcada et al. 2004; Franciosini et al. 2006; Caballero et al.
2007)\footnote{All the {\em ROSAT} sources associated by Mokler \& Stelzer
(2002) to candidate brown dwarfs in $\sigma$~Orionis seem to have their origin
in more massive cluster members at several arcsec (e.g. \object{S\,Ori
J053948.1--022914} + \object{B 1.01--319}; Flesch \& Hardcastle 2004; Burningham
et al. 2005).}. 
The possibility of Mayrit 53049 and Mayrit 91024 possessing discs has been
also considered by Hern\'andez et al. (2007).

Accounting for the 13 X-ray sources with optical and near-infrared counterpart
from Franciosini et al. (2006) (Table \ref{xray}) and the ten new X-ray sources
detected in the {\em Chandra} ACIS and/or HRC-I images (Table \ref{newxray}),
there are 23 young stars with X-ray emission.
The non-emitters are Mayrit 11238 ($\sigma$~Ori C -- an A0V star without any
strong stellar wind or any corona), Mayrit 123000
(P053844--0233 -- maybe a late G-type star), Mayrit 91024 (with a probable disc)
and two very faint stars close to the substellar boundary 
with expected very faint X-ray emission (Mayrit 89175 and Mayrit 50279;
presented here for the first time).
The frequency of X-ray emitters in the centre of the $\sigma$~Orionis cluster
with masses between $\sim$20\,M$_\odot$ (the O9.5V) and the
hydrogen-burning limit is as high as 80$\pm$20\,\%.
In the magnitude interval 11.2\,mag $< I <$ 15.4\,mag there is only one star
without X-ray emission, Mayrit 91024.
Therefore, searches for young objects in clusters using exclusively X ray data
could be efficient and have a relatively large success rate.

\subsection{Spatial density}
\label{spatial}

The presence of 27 star cluster members and candidates in an area of only
$\sim$17\,arcmin$^2$ leads to a maximum stellar surface spatial density as a
large as 1.6$\pm$0.1\,arcmin$^{-2}$, larger than previously computed.
If the 27 stars were cluster members and would be located within a cube of
$\sim$4.1\,arcmin size centred in $\sigma$~Ori AB ($\sim$0.43\,pc at the most
probable heliocentric distance to the cluster of 360\,pc), then the stellar
volumetric density would be 340$\pm$70\,pc$^{-3}$. 
On the one hand, if the cluster would be located further, then the density would
be even larger. 
On the other hand, due to a projection effect, not all the stars are within the
cube.
However, due to the intense gravitational well close to the cluster centre, most
of them are probably within the volume. 
The density can not be in any case less than $\sim$100\,pc$^{-3}$ (one third of
the stars contained in the cube).
This value should be compared with the central stellar density proposed by
Sherry et al. (2004), of only $\sim$3\,pc$^{-3}$ (although they studied only the
0.2--1.0\,M$_\odot$ mass interval and noted that several parameter combinations
fit well the spatial distribution to a simple King model, with spatial densities
ranging from 2.5 to 20\,pc$^{-3}$).

The value of the stellar surface density in the area, $\rho$ =
1.6$\pm$0.1\,arcmin$^{-2}$, is 6.2 times larger than the central surface density
of substellar objects in $\sigma$~Orionis from the exponential fit to the radial
distribution by B\'ejar et al. (2004a). 
From their work, it is expected to be $\sim$4 substellar objects in this
optical/near-infrared survey area, while I have found none.
However, it {\em does not} imply a lack of substellar objects in the cluster
centre, since this survey is only sensitive to the detection of brown dwarfs
brighter than $J \sim$ 16.5\,mag. 
In particular, it is expected to be only $\sim$1 brown dwarf in the area between
this magnitude and the substellar boundary from the data of B\'ejar et al.
(2004a). 
Nonetheless, more work is needed to study the variation of the brown
dwarf-to-star ratio with the separation to the cluster centre.

\subsection{Additional remarkable objects}

\begin{table*}
\centering
\caption{Remarkable objects not shown in Table \ref{photon} and with
near-infrared colours of field dwarfs with L spectral type.}
\label{remarkable}
\begin{tabular}{l c c c c c c c}
\hline
Identification 		& $\alpha$ 		& $\delta$		& $I$	 	& $J$ 		 & $H$		  & $K_{\rm s}$	    & Name 	\\
 			& (J2000) 		& (J2000)		& (mag)	 	& (mag)	 	 & (mag)	  & (mag)	    & 		\\
\hline
\object{Mayrit 72345}	& 05 38 43.5 		& $-$02 34 50		& $>$18.5	& 18.98$\pm$0.15 & 17.93$\pm$0.13 & 17.18$\pm$0.13  & NX 77	\\ %
\object{Mayrit 111335}	& 05 38 41.9 		& $-$02 34 29		& $>$18.5	& 18.78$\pm$0.12 & 17.69$\pm$0.09 & 16.91$\pm$0.11  & 		\\ %
\hline
\end{tabular}
\end{table*}

\subsubsection{Mayrit 92149 AB (R053847--0237)}

Mayrit 92149 was classified by Wolk (1996) as a K5-type star with X-ray
emission. 
He measured the pseudo-equivalent widths of H$\alpha$ and Li {\sc i} at --9.56
and +0.26\,\AA.
It is an accreting classical T Tauri star according to the  empirical criterion
of Barrado y Navascu\'es \& Mart\'{\i}n (2003).
The X-ray emission was afterwards confirmed by Franciosini et al. (2006).
Sherry et al. (2004) catalogued two objects separated 0.75\,arcsec and with
roughly the same $VRI_CJHK_{\rm s}$ magnitudes (SWW 102 and SWW 149) in the 
coordinates of Mayrit 92149.

I have confirmed that Mayrit 92149 is a visual binary.
Both components are separated by $\rho \sim$1.9\,arcsec, about
700\,AU at the cluster heliocentric distance ($\theta \sim$ 60\,deg).
I have also measured the difference of magnitudes between both of them in the
IAC-80 $VRI$ short exposures and in the CAIN-II near-infrared images:
$\Delta V$ = 1.1$\pm$0.4\,mag, $\Delta R$ = 0.80$\pm$0.17\,mag, $\Delta I$ =
0.62$\pm$0.10\,mag, $\Delta J \sim$ 0.6\,mag, $\Delta H$ = 0.57$\pm$0.02\,mag,
$\Delta K_{\rm s}$ = 0.88$\pm$0.08\,mag. 
The difference of magnitude in the reddest pass-band indicates that the
brightest component has a $K_{\rm s}$ flux excess.
This may be related to a disc, which would also be responsible of the strong
H$\alpha$ emission.
Furthermore, the composite (Mayrit 92149 AB) 2MASS $J-K_{\rm s}$ colour is
1.13$\pm$0.16\,mag, as red within the uncertainties as the least red star with
disc in the survey area (Section \ref{discs}).
The secondary in the system, although not having a disc, satisfies the
photometric criterion of membership in $\sigma$~Orionis.

To date, there were known three types of binaries in the cluster: 
$(i)$ spectroscopic binaries, like \object{OriNTT 429} and \object{S\,Ori 36}
(Lee et al. 1994; Kenyon et al. 2005);
$(ii)$ tight binaries with $\rho \sim$ 0.3--0.4\,arcsec, only resolvable with
micrometer, speckle, adaptive optics or lucky imaging, like $\sigma$~Ori AB,
\object{HD 37525} AB and \object{[W96] 4771--899} AB (Heintz et al. 1997;
Caballero 2005; R. Rebolo, priv. comm.); and
$(iii)$ wide binaries separated between 3.3 and 10\,arcsec, like $\sigma$~Ori
AB + IRS1, [W96] 4771--899 AB + \object{S\,Ori J053847.7--022711} and
\object{[SE2004] 70} + \object{S\,Ori 68} (van Loon \& Oliveira 2003; Caballero
2005; Caballero et al. 2006).
Mayrit 92149 AB is among the first binaries found in the cluster
with an intermediate separation between tight and wide binaries.

There is other binary in the area, claimed by Wolk (1996), Mayrit 102101.
It may be a spectroscopic binary because there is no hint for multiplicity in my
images.

\subsubsection{Mayrit 21023 and Mayrit 30241}
\label{MM}

\begin{figure}
\includegraphics[width=0.47\textwidth]{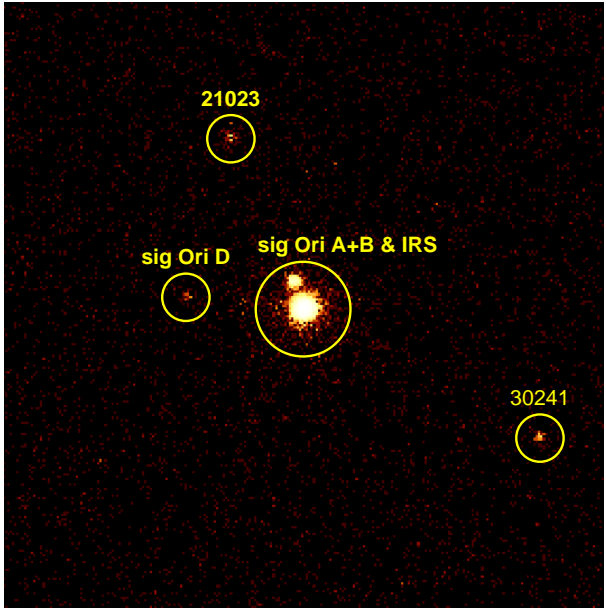}
\caption{Zoom of the HRI-C/{\em Chandra} image in the central 1 $\times$
1\,arcmin$^2$. 
Four objects are labeled and marked with circles.
$\sigma$~Ori~D, Mayrit 21023 and the 2MASS source Mayrit 30241 are X-ray
emitters reported here for the first time.
The system $\sigma$~Ori AB + IRS1 is also resolved in this image.}
\label{F04}
\end{figure}

In the central 1 $\times$ 1\,arcmin$^2$ (see Fig. \ref{F04}), there are three
previously unknown X-ray sources.
In that region, the two faintest X-ray stars in the optical are Mayrit 21023 and
Mayrit 30241 (the third X-ray source is $\sigma$~Ori D). 
The latter may have gone unnoticed due to its closeness to the central
O9.5V + B0.5V pair ($\rho \approx$ 30\,arcsec), although its relative brightness
($I$ = 13.14$\pm$0.09\,mag).
As discussed in Section \ref{discs}, it has a $K_{\rm s}$ excess that informs on
a surrounding disc.

The same scenario may also be applied to Mayrit 21023, which photometry is
however more contaminated due to the O9.5V+B0.5V pair.
This contamination is very important in the $I$ band (see Table \ref{photon}).
This star was firstly identified by Caballero (2005) with the NAOMI+INGRID
adaptive optics system at the 4.2\,m William Herschel Telescope.
Afterwards, Sherry et al. (2005) classified Mayrit 21023 as a very red
foreground M dwarf based on non-reliable $I$-band photometry.
According to them, Mayrit 21023 would be brighter than Mayrit 114305, which is
the brightest star in the field after the components of the $\sigma$~Ori system.
Besides, Mayrit 21023 would have an extraordinary peculiar $I-J$ colour of
--2.55\,mag (all main-sequence stars, giants and brown dwarfs with spectral
types between O9 and T8 have $I-J$ colours redder than --1.0\,mag).
On the contrary, I consider it a normal cluster member which requires further
spectro-photometric analysis in the optical.

\subsubsection{2MASS J05384652--0235479 and 2MASS J05384454--0235349}

The source \object{2MASS J05384652--0235479} is a candidate cluster member
proposed by Sherry et al. (2005) ({\em star 1}).
They tabulated its $V$ magnitude at 14.8$\pm$0.2\,mag, and considered it as a
cluster member with a mass of $\sim$0.7\,M$_\odot$.
However, its $I$ magnitude does not match again with those catalogued by
DENIS or derived from the deep $I$-band IAC-80 image.
If it were a cluster member, 2MASS J05384652--0235479 would be a peculiar young
star with ~ $V-I$ = 0.0$\pm$0.3\,mag, $I-J$ = 0.1$\pm$0.2\,mag and $J-K_{\rm
s}$ = 0.38$\pm$0.18\,mag. 
The near-infrared photometry is from 2MASS because the TCS images at 30\,arcsec
to the north-east to $\sigma$~Ori AB are highly affected by its glare. 
The ~ location ~ of ~ 2MASS J05384652--0235479, in between $\sigma$~Ori AB and
$\sigma$~Ori E, makes difficult photometric and spectroscopic follow-ups
to ascertain its nature.

Finally, 2MASS catalogued another source in the glare of $\sigma$~Ori,
\object{2MASS J05384454--0235349}. 
Its $I-K_{\rm s}$ colour, larger than 2.0\,mag, is not restrictive.
However, its blue $J-K_{\rm s}$ colour, of --0.12$\pm$0.16\,mag, for its $K_{\rm
s}$ magnitude, of 15.83$\pm$0.10\,mag, suggests that it may be an A--F star or a
galaxy in the background rather than a cluster member with abnormal colours.

\subsubsection{Mayrit 72345 (NX 77) and Mayrit 111335}

There are five near-infrared sources fainter than $J$ = 18.5\,mag without
optical counterpart and with $J-K_{\rm s}$ colours redder than 1.5\,mag.
Only two out of them have $JHK_{\rm s}$ colours similar to those of intermediate
and late L-type field dwarfs (e.g. Dahn et al. 2002).
One of these near-infrared sources, preliminary named Mayrit 72345, is
located at 5.2\,arcsec to the X-ray source NX 77 (found by Franciosini et
al. 2006, but {\em not} identified in the HRC-I/{\em Chandra} images).
This angular separation is typical for very faint X-ray sources in the 
EPIC/{\em XMM-Newton} images (e.g. 5.0\,arcsec for \object{SWW~203}, 4.9\,arcsec
for \object{S\,Ori~6}; Franciosini et al. 2006).
There is no other brighter near-infrared source at less than 30\,arcsec from the
X-ray source.
Therefore, Mayrit 72345 is likely the near-infrared counterpart of NX 77.
This fact, if confirmed, is of extreme importance: $\sigma$~Orionis members with
the same $J$-band magnitude, $\sim$19.0\,mag, have intermediate L-type spectral
types and most-probable masses in the planetary domain, at about
0.009--0.007\,M$_\odot$.
Mayrit 72345 might be, therefore, an X-ray source one order of
magnitude less massive than the stars just above the substellar boundary.
It is obvious that further investigation is needed to assess its membership in
the cluster.
Its coordinates and magnitudes are given in Table \ref{remarkable}.
The same values of the other likely L-type object in $\sigma$
Orionis, also hypothetically in the planetary domain but without X-ray emission,
are provided as well (Mayrit 111335). 
Both of them follow the spectro-photometric cluster sequence.
The probability of they being field dwarf contaminants in a survey
of only 17\,arcmin$^2$ is extremely low.
From the detailed study of contamination by ultracool dwarfs in
$\sigma$~Orionis by Caballero et al. (2007), it is expected that only 0.04
L-type field dwarfs contaminate this survey.
The other three near-infrared sources with $J-K_{\rm s} >$ 1.5\,mag are probably
distant red galaxies and are not catalogued here.

\section{Conclusions}

An optical/near-infrared survey in the centre of the $\sigma$~Orionis cluster is
presented.
I have covered about 4 $\times$ 4\,arcmin$^2$ centred on the $\sigma$~Ori
multiple star system in the $IJHK_{\rm s}$ bands.
The near-infrared images reach to the magnitudes of young planetary-mass objects
in the cluster.
The survey has been complemented with spectroscopic information from the
literature, published X-ray data and a new analysis of public data of the {\em
Chandra} Space Telescope.

I have selected 28 candidate cluster members from their position in the $I$ vs.
$I-K_{\rm s}$ colour-magnitude diagram.
Eleven display spectroscopic features of youth and thirteen were known to be
X-ray emitters.
I have found the X-ray counterparts of another ten sources, previously
unknown.  
Six out of the 28 candidate cluster members are firstly found in this work.
They are a $\sim$0.6\,M$_\odot$ star, four very low mass stars and a very
red object, named Mayrit 111208.
The latter, apart of it being an X-ray emitter, displays the reddest $J-K_{\rm
s}$ colour among known $\sigma$~Orionis cluster ~ members ~ ($J-K_{\rm s}$ =
2.22$\pm$0.13\,mag).
It may be a classical T Tauri low-mass star with an extended,
probably edge-on, disc.
Other three low-mass stars show $K_{\rm s}$ excess.

The stellar spatial density, of up to 340$\pm$70\,pc$^{-3}$, is several orders
of magnitude larger than previously estimated.
I have not found a lack of substellar objects in the cluster centre.
The frequency of X-ray emitters in the region is also very high, of
80$\pm$20\,\%. 

Finally, I have reported on several remarkable objects: a binary with a
projected physical separation of $\sim$700\,AU (among the first in its
class in $\sigma$~Orionis), two cluster stars with X-ray emission located at
20--30\,arcsec from the O9.5V + B0.5V central system (they are closer to
$\sigma$ Ori AB than $\sigma$~Ori E), two 2MASS sources with peculiar colours
and two objects with the near-infrared magnitudes and colours typical of known
L-type objects in the cluster (one of them is likely associated to an X-ray
source).

\acknowledgements

This work is partially based in the Chapter~4 of my PhD thesis (Caballero 2006,
Universidad de La Laguna; language: Spanish; printing: 25 copies).
Therefore, I am deeply in debt to my thesis supervisors, R. Rebolo and V. J.
S\'anchez B\'ejar, to the internal referee, E. L. Mart\'{\i}n, and to the
board of examiners of my thesis. 
I also thank J. Sanz Forcada for very helpful comments.
The Telescopio IAC-80 and the Telescopio Carlos S\'anchez are
operated on the island of Tenerife by the Instituto de Astrof\'{\i}sica
de Canarias in the Spanish Observatorio del Teide of the Instituto de 
Astrof\'{\i}sica de Canarias.	
This publication makes use of data products from the Two Micron All Sky
Survey, which is a joint project of the University of Massachusetts and
the Infrared Processing and Analysis Center/California Institute of
Technology, funded by the National Aeronautics and Space
Administration and the National Science Foundation.
IRAF is distributed by National Optical Astronomy Observatories,
which are operated by the Association of Universities for Research in
Astronomy, Inc., under cooperative agreement with the National Science
Foundation.
This research has made use of the SIMBAD database, operated at CDS,
Strasbourg, France.

\end{document}